\documentclass{elsart}

\usepackage{natbib}

\usepackage{epsfig}

\usepackage{amssymb}
\begin{document}

\begin{frontmatter}

% Title, authors and addresses

% use the thanksref command within \title, \author or \address for footnotes;
% use the corauthref command within \author for corresponding author footnotes;
% use the ead command for the email address,
% and the form \ead[url] for the home page:
% \title{Title\thanksref{label1}}
% \thanks[label1]{}
% \author{Name\corauthref{cor1}\thanksref{label2}}
% \ead{email address}
% \ead[url]{home page}
% \thanks[label2]{}
% \corauth[cor1]{}
% \address{Address\thanksref{label3}}
% \thanks[label3]{}

\title{Mission Design for the \\ Laser Astrometric Test Of Relativity}

% use optional labels to link authors explicitly to addresses:
% \author[label1,label2]{}
% \address[label1]{}
% \address[label2]{}

\author[label1]{Slava G. Turyshev\corauthref{cor1}},
\corauth[cor1]{Corresponding author.}
\ead{turyshev@jpl.nasa.gov}
\author[label1]{Michael Shao},
\author[label2]{Kenneth Nordtvedt, Jr.}
\address[label1]{NASA Jet Propulsion Laboratory
4800 Oak Grove Drive, Pasadena, CA 91109}

\address[label2]{Northwest Analysis, 118 Sourdough Ridge Road, Bozeman, MT 59715 U.S.A.}

\begin{abstract}
% Text of abstract

This paper focuses on the mission design for the Laser Astrometric Test Of Relativity (LATOR). This mission uses laser interferometry between two micro-spacecraft whose lines of sight pass close by the Sun to accurately measure deflection of light in the solar gravity.  The key element of the experimental design is a redundant geometry optical truss provided by a long-baseline (100 m) multi-channel stellar optical interferometer placed on the International Space Station (ISS). The spatial interferometer is used for measuring the angles between the two spacecraft and for orbit determination purposes. The geometric redundancy enables LATOR to measure the departure from Euclidean geometry caused by the solar gravity field to a very high accuracy.
Such a design enables LATOR to improve the value of the parameterized post-Newtonian (PPN) parameter $\gamma$ to unprecedented levels of accuracy of 1 part in 10$^{8}$; the misison will also measure effects of the next post-Newtonian order ($\propto G^2$) of light deflection resulting from gravity's intrinsic non-linearity.  The solar quadrupole moment parameter, $J_2$, will be measured with high precision, as well as a variety of other relativistic effects including Lense-Thirring precession.  LATOR will lead to very robust advances in the tests of Fundamental physics: this mission could discover a violation or extension of general relativity, or reveal the presence of an additional long range interaction in the physical law.  
 
\end{abstract}

\begin{keyword}
% keywords here, in the form: keyword \sep keyword
tests of general relativity \sep astrometry \sep scalar-tensor theories \sep laser ranging
% PACS codes here, in the form: 
\PACS 04.80.-y \sep 95.10.Eg \sep 95.55.Pe   

\end{keyword}

\end{frontmatter}

% main text
%*********************1) INTRODUCTION
\section{Introduction}

Following its successful confirmation by the 1919 Eddington's expedition, Einstein's general theory of relativity (GR) has been verified at ever higher accuracy. Thus, microwave ranging to the Viking Lander on Mars yielded accuracy  $\sim$0.1\% in the tests of GR \citep{viking_shapiro1,viking_reasen}. The astrometric observations of quasars on the solar background performed with Very-Long Baseline Interferometry (VLBI) improved the accuracy of the tests of gravity to $\sim$0.04\% \citep{RoberstonCarter91,Shapiro_SS_etal_2004}. 
Lunar Laser Ranging (LLR),  a continuing legacy of the Apollo program, provided $\sim$0.01\% verification of GR via precision measurements of the lunar orbit \citep{Ken_LLR30years99,JimSkipJean96,pr01}. Finally, the recent experiments with the Cassini spacecraft improved the accuracy of the tests to $\sim$0.0023\% \citep{cassini_ber}. As a result, GR became the standard theory of gravity when astrometry and spacecraft navigation are concerned. 

However, the continued inability to merge gravity with quantum mechanics, and recent cosmological observations indicate that the pure tensor gravity of GR needs modification.  Progress in scalar-tensor extensions of gravity which are consistent with present cosmological models \citep{damour_nordtvedt1,DPV02,Damour_EFarese96} motivate new searches for very small deviations of relativistic gravity in the solar system, at levels of 10$^{-5}$ to 10$^{-7}$ of the post-Newtonian effects or essentially to achieve accuracy that is compatible to the size of the effects of the second order in the gravitational field strength ($\propto G^2$).  This will require a several order of magnitude improvement in experimental precision from present tests. The ability to measure the first order light deflection term at the accuracy comparable with the effects of the second order is of the utmost importance for the gravitational theory and is the challenge for the 21st century fundamental physics. 

When the light deflection in solar gravity is concerned, the magnitude of the first order effect as predicted by GR for the light ray just grazing the limb of the Sun is $\sim1.75$ arcsecond. (Note that 1 arcsecond $\simeq5~\mu$rad; when convenient, below we will use the units of radians and arcseconds interchangeably.) The effect varies inversely with the impact parameter. The second order term is almost six orders of magnitude smaller resulting in  $\sim 3.5$ microarcseconds ($\mu$as) light deflection effect, and which falls off inversely as the square of the light ray's impact parameter \citep{Ken_2PPN_87,lator_cqg2004}. The smallness of the effects emphasize the fact that, among the four forces of nature, gravity is the weakest interaction; it acts at very long distances and controls the large-scale structure of the universe, thus, making the precision tests of gravity a very challenging task. 

The LATOR  mission is designed to directly address the challenges discussed above \citep{lator_cqg2004}. The test will be performed in the solar gravity field using optical interferometry between two micro-spacecraft.  Precise measurements of the angular position of the spacecraft will be made using a fiber coupled multi-chanelled optical interferometer on the ISS with a 100 m baseline. The primary objective of LATOR will be to measure the gravitational deflection of light by the solar gravity to accuracy of 0.1 picoradians (prad) ($\sim0.02 ~\mu$as). 

In conjunction with laser ranging among the spacecraft and the ISS, LATOR will allow measurements of the gravitational deflection by a factor of more than 3,000 better than had recently been accomplished with the Cassini spacecraft. In particular, this mission will not only measure the key Eddington parameter $\gamma$ to unprecedented levels of accuracy of one part in 10$^8$. The Eddington parameter $\gamma$, whose value in GR is unity, is perhaps the most fundamental PPN parameter, in that $(1-\gamma)$ is a measure, for example, of the fractional strength of the scalar gravity interaction in scalar-tensor theories of gravity \citep{damour_nordtvedt1,Damour_EFarese96,lator_cqg2004}.  Within perturbation theory for such theories, all other PPN parameters to all relativistic orders collapse to their GR values in proportion to $(1-\gamma)$. This is why measurement of the first order light deflection effect at the level of accuracy comparable with the second-order contribution would provide the crucial information separating alternative scalar-tensor theories of gravity from GR \citep{Ken_2PPN_87} and also to probe possible ways for gravity quantization and to test modern theories of cosmological evolution discussed in the previous section.  LATOR is designed to directly address this issue with an unprecedented accuracy; it will also reach ability to measure the next post-Newtonian order ($\propto G^2$) of light deflection with accuracy to 1 part in $10^3$.

The LATOR mission technologically is a very sound concept; all technologies that are needed for its success have been already demonstrated as a part of the JPL's Space Interferometry Mission (SIM) development \citep{lator_cqg2004}.   The LATOR concept arose from several developments at JPL that initially enabled optical astrometry and metrology, and also led to developing expertize needed for the precision gravity experiments. Technology that has become available in the last several years such as low cost microspacecraft, medium power highly efficient solid state and fiber lasers, and the development of long range interferometric techniques make possible an unprecedented factor of 3,000 improvement in this test of GR possible. This mission is unique and is the natural next step in solar system gravity experiments which fully exploits modern technologies.

This paper organized as follows: Section \ref{sec:lator_description} provides an overview for the LATOR experiment including the current mission design.  Section \ref{sec:error_bud} discusses a preliminary  analysis of the expected performance of various subsystems. We conclude with a discussion of future developments for LATOR. 

\section{Overview of LATOR}
\label{sec:lator_description}

\begin{figure*}[t!]
 \begin{center}
\noindent    
\psfig{figure=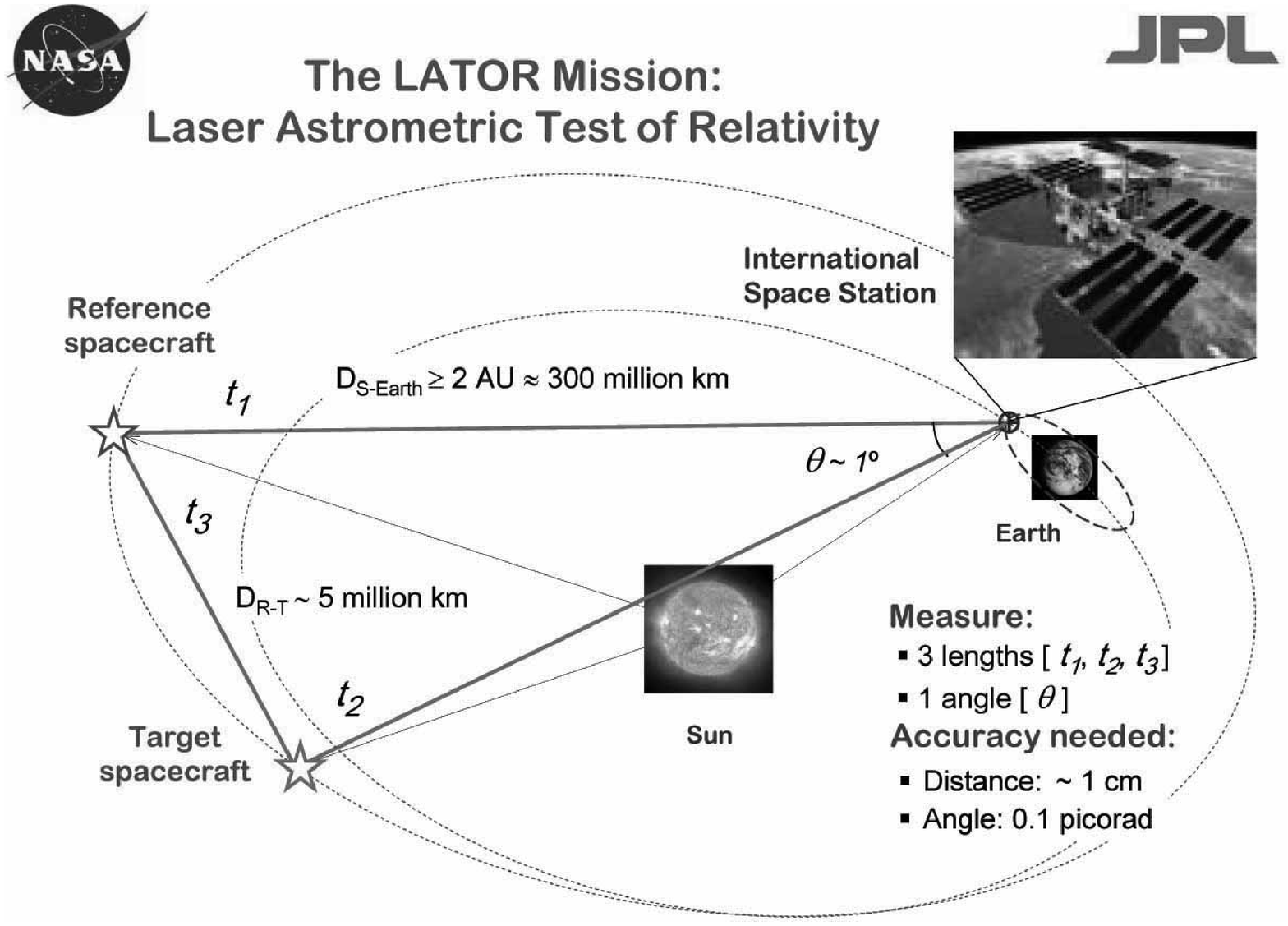,width=140mm}%,height=90mm}
\end{center}
\vskip -10pt 
  \caption{The overall geometry of the LATOR experiment.  
 \label{fig:lator}}
\end{figure*}

The LATOR experiment uses laser interferometry between two micro-space\-craft (placed in heliocentric orbits, at distances $\sim$ 1 AU from the Sun) whose lines of sight pass close by the Sun to accurately measure deflection of light in the solar gravity. Another component of the experimental design is a long-baseline ($\sim$100 m) multi-channel stellar optical interferometer placed on the ISS. Figure \ref{fig:lator} shows the general concept for the LATOR missions including the mission-related geometry, experiment details  and required accuracies. 

%************

%**************

\subsection{LATOR Mission Design}

 The schematic of the LATOR experiment is quite simple (see Fig.~\ref{fig:lator}). Two spacecraft are injected into a solar orbit on the opposite side of the Sun from the Earth. Each spacecraft transmits a laser beam which is detected by a long baseline ($\sim$ 100 m) optical interferometer on the ISS. This interferometer measures the apparent angle between the two spacecraft. In addition, each spacecraft contains laser ranging systems which measure the arms of the triangle formed by the two spacecraft and the ISS. From these measurements, the angle between the two spacecraft viewed from the ISS can be computed using Euclidean geometry. In the absence of gravity, this computed angle will be identical to the apparent angle measured by the interferometer. The difference between the two angles is a measure of the deflection of light by gravity.

As a baseline design for the LATOR orbit, both spacecraft will be launched on the same launch vehicle. Almost immediately after the launch there will be a 30 m/s maneuver that separates the two spacecraft on their 3:2 Earth resonant orbits \citep{lator_cqg2004}.  The sequence of events that occurs during each observation period will be initiated at the beginning of each orbit of the ISS. It assumed that bore sighting of the spacecraft attitude with the spacecraft transmitters and receivers have already been accomplished. This sequence of operations is focused on establishing the ISS to spacecraft link. The interspacecraft link is assumed to be continuously established after deployment, since the spacecraft never lose line of sight with one another. 

As evident from Figure \ref{fig:lator}, the key element of the LATOR experiment is a redundant geometry optical truss to measure the departure from Euclidean geometry caused by gravity.  The triangle in figure has three independent quantities but three arms are monitored with laser metrology. From three measurements one can calculate the Euclidean value for any angle in this triangle.  In Euclidean geometry these measurements should agree to high accuracy.  This geometric redundancy enables LATOR to measure the departure from Euclidean geometry caused by the solar gravity field to a very high accuracy. The difference in the measured angle and its Euclidean value is the non-Euclidean signal.  

The shortening of the interferometric baseline (as compared to the previously studied version \citep{yu94}) is achieved solely by going into space to avoid the atmospheric turbulence and Earth's seismic vibrations. On the space station, all vibrations can be made common mode for both ends of the interferometer by coupling them by an external laser truss. This relaxes the constraint on the separation between the spacecraft, allowing it to be as large as few degrees, as seen from the Earth. Additionally, the orbital motion of the ISS provides variability in the interferometer's baseline projection as needed to resolve the fringe ambiguity of the stable laser light detection by an interferometer \citep{lator_cqg2004}.

\subsection{Trajectory -- a 3:2 Earth Resonant Orbit}

In this section we outline the basic elements of the LATOR trajectory design.
The objective of the LATOR mission includes placing two spacecraft into a heliocentric orbit with a one year period so that observations may be made when the spacecraft are behind the Sun as viewed from the ISS.  The observations involve the measurement of distance of the two spacecraft using an interferometer on-board the ISS to determine bending of light by the Sun.  The two spacecraft are to be separated by about 1$^\circ$, as viewed from the ISS.  
With the help of the JPL Advanced Project Design Team (Team X), we recently conducted a detailed mission design studies \citep{teamx}. In particular, we analyzed various trajectory options for the deep-space flight segment of LATOR, using both Orbit Determination Program (ODP) and Satellite Orbit Analysis Program (SOAP) --  the two standard JPL navigation software packages. 

An excellent choice for the LATOR orbit was found when we studied  a possibility of launching spacecraft into the orbit with a 3:2 resonance  with the Earth \citep{teamx}. (The 3:2 resonance occurs when the Earth does 3 revolutions around the Sun while the spacecraft does exactly 2 revolutions of a 1.5 year orbit. The exact period of the orbit may vary slightly ($<$1\%) from a 3:2 resonance depending on the time of launch.) For this orbit, in 13 months after the launch, the spacecraft are within $\sim10^\circ$ of the Sun with first occultation occuring in 15 months after launch \citep{lator_cqg2004}.  At this point, LATOR is orbiting at a slower speed than the Earth, but as LATOR approaches its perihelion, its motion in the sky begins to reverse and the spacecraft is again occulted by the Sun 18 months after launch.  As the spacecraft slows down and moves out toward aphelion, its motion in the sky reverses again and it is occulted by the Sun for the third and final time 21 months after launch.  This entire process will again repeat itself in about 3 years after the initial occultation, however, there may be a small maneuver required to allow for more occultations.

The 3:2 Earth resonant orbit provides an almost ideal trajectory for the LATOR spacecraft, specifically i) it imposes no restrictions on the time of launch; ii) with a small propulsion maneuver after the launch, it places the two LATOR spacecraft at the distance of $\leq3.5^\circ$ ($\sim14~R\odot$) for the entire duration of the experiment ($\sim$8 months); iii) it provides three solar conjunctions even during the nominal mission lifetime of 22 months, all within a 7 month period; iv) at a cost of an small additional maneuver, it offers a possibility of achieving  small orbital inclinations (to enable measurements at different solar latitudes), and, finally, v) this orbit offers a very slow change in the Sun-Earth-Probe (SEP) angle of $\sim 1R_\odot$ in 4 days. As such, this orbit represents a very attractive choice for  LATOR. In particular, there is an option to have the two spacecraft move in  opposite directions during the solar conjunctions. This option will increase the amount of $\Delta v$ LATOR should carry on-board, but it significantly reduces the experiment's dependence on the accuracy of determination of the solar impact parameter. This particular option is currently being investigated and results will be reported elsewhere. 

\subsection{Optical Design}
\label{sec:opt_design}

In this Section we consider the basic elements of the LATOR optical design. 

A single aperture interferometer on the ISS consists of three 20 cm diameter telescopes. One of the telescopes with a very narrow bandwidth  filter in front and with an InGAs camera at its focal plane, sensitive to the 1.3 $\mu$m laser light, serves as the acquisition telescope to locate the spacecraft near the Sun.

The second telescope emits the directing beacon to the spacecraft. Both spacecraft are served out of one telescope by a pair of piezo controlled mirrors placed on the focal plane. The properly collimated laser light ($\sim$10W) is injected into the telescope focal plane and deflected in the right direction by the piezo-actuated mirrors. 

The third telescope is the laser light tracking interferometer input aperture which can track both spacecraft at the same time. To eliminate beam walk on the critical elements of this telescope, two piezo-electric X-Y-Z stages are used to move two single-mode fiber tips on a spherical surface while maintaining focus and beam position on the fibers and other optics. Dithering at a few Hz is used to make the alignment to the fibers and the subsequent tracking of the two spacecraft completely automatic. The interferometric tracking telescopes are coupled together by a network of single-mode fibers whose relative length changes are measured internally by a heterodyne metrology system to an accuracy of less than 10 pm.

%************
\begin{figure*}[!t!]
 \begin{center}
\noindent    
\psfig{figure=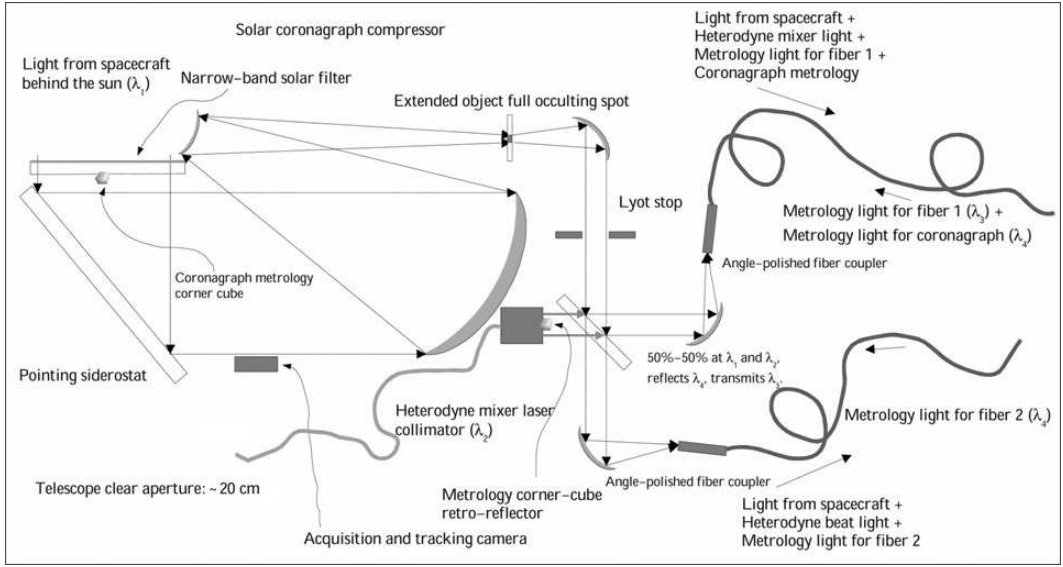,width=140mm}%,height=90mm}
\end{center}
\vskip -0pt 
  \caption{Basic elements of the LATOR optical design. 
The laser light (together with the solar background) falls onto a full aperture ($\sim 20$cm) narrow band-pass filter with $\sim 10^{-4}$ suppression capabilities and also illuminates the baseline metrology corner cube.  After that, the remaining light is falling onto a steering flat mirror where it will be reflected to an off-axis telescope with no central obscuration (needed for metrology). This is when it enters the solar coronograph compressor by first going through a 1/2 plane focal plane occulter and then coming to a Lyot stop. At the Lyot stop, the background solar light is reduced by a factor of $10^{6}$. The combination of a narrow band-pass filter and coronograph enables the solar luminosity reduction from $V=-26$ to $V=4$ (as measured at the ISS), thus, enabling the LATOR precision observations.
\label{fig:optical_design}}
\end{figure*} 
%**************

The spacecraft  are identical in construction and contain a relatively high powered (2 W), stable (2 MHz per hour $\sim$  500 Hz per second), small cavity fiber-amplified laser at 1.3 $\mu$m. Three quarters of the power of this laser is pointed to the Earth through a 20 cm aperture telescope and its phase is tracked by the interferometer. With the available power and the beam divergence, there are enough photons to track the slowly drifting phase of the laser light. The remaining part of the laser power is diverted to another telescope, which points towards the other spacecraft. In addition to the two transmitting telescopes, each spacecraft has two receiving telescopes.  The receiving telescope on the ISS, which points toward the area near the Sun, has laser line filters and a simple knife-edge coronagraph to suppress the Sun light to 1 part in $10^4$ of the light level of the light received from the space station (see Figure \ref{fig:optical_design} for a conceptual design). The receiving telescope that points to the other spacecraft is free of the Sun light filter and the coronagraph.

In order to have adequate rejection of the solar background surrounding the laser uplink the ISS and spacecraft optical systems must include a coronagraph.  A 20~cm telescope forms an image on the chronographic stop. This stop consists of a knife-edge mask placed 6 arcseconds beyond the solar limb. The transmitted light is then reimaged onto a Lyot stop, which transmits 88\% of the incident intensity. Finally, the light is
reimaged onto the tracking detector. Note that, in addition a combination of a wideband interference filter and a narrow band Faraday anomalous dispersion optical filter (FADOF) \citep{lator_cqg2004} will be used to reject light outside a 0.005 nm band around the laser line. Thus, for ISS-SC receiver/transmitter, the incoming signal will subdivided with one portion going to a high bandwidth detector and the other to an acquisition and tracking CCD array (see Fig.~\ref{fig:optical_design}). Using a $64 \times 64$ CCD array with pixels sized to a diffraction limited spot, this array will have a 5 arcmin field of view which is greater than the pointing knowledge of the attitude control system and the point ahead angle (40 arcsec). After acquisition of the ISS beacon, a $2\times 2$ element subarray of the CCD will be used as a quad cell to control the ISS-SC two axis steering mirror. This pointing mirror is common to both the receiver and transmitter channel to minimize misalignments between the two optical systems due to thermal variations. The pointing mirror will have 10 arcminute throw and a pointing accuracy of 0.5 arcsec which will enable placement of the uplink signal on the high bandwidth detector. Similar design elements will be implemented in the other optical packages.

Our preliminary analysis indicates that LATOR will achieve a significant stray light rejection, even observing at the solar limb. In fact, the flux from the solar surface may be minimized by a factor of $10^4$. In addition to decreasing the stray solar radiation, the coronograph will decrease the transmission of the laser signal by 78\% (for a signal 12 arcsec from limb) due to coronographic transmission and broadening of the point spread function. At these levels of solar rejection, it is possible for the spectral filter to reject enough starlight to acquire the laser beacon (even at the $\sim$~fW level). It is interesting that without the coronograph, the stray light from the Sun, decreases proportionally to the distance from the limb, but with the use of the coronograph, it decreases as a square of the distance from the limb.  

In addition to the four telescopes they carry, the spacecraft also carry a tiny (2.5 cm) telescope with a CCD camera. This telescope is used to initially point the spacecraft directly towards the Sun so that their signal may be seen at the space station. One more of these small telescopes may also be installed at right angles to the first one to determine the spacecraft attitude using known, bright stars. The receiving telescope looking towards the other spacecraft may be used for this purpose part of the time, reducing hardware complexity. Star trackers with this construction have been demonstrated many years ago and they are readily available. A small RF transponder with an omni-directional antenna is also included in the instrument package to track the s/c while they are on their way to assume the orbital position needed for the experiment. 

The LATOR experiment has a number of advantages over techniques which use radio waves to measure gravitational light deflection. Advances in optical communications technology, allow low bandwidth telecommunications with the LATOR spacecraft without having to deploy high gain radio antennae needed to communicate through the solar corona. The use of the monochromatic light enables the observation of the spacecraft almost at the limb of the Sun, as seen from the ISS. The use of narrowband filters, coronagraph optics and heterodyne detection will suppress background light to a level where the solar background is no longer the dominant noise source. In addition, the short wavelength allows much more efficient links with smaller apertures, thereby eliminating the need for a deployable antenna. Finally, the use of the ISS will allow conducting the test above the Earth's atmosphere -- the major source of astrometric noise for any ground based interferometer. This fact justifies LATOR as a space mission.

\section{LATOR Expected Performance}
\label{sec:error_bud}

The goal of measuring deflection of light in solar gravity with accuracy of one part in $10^{8}$ requires serious consideration of systematic errors. This work requires a significant effort to properly identify the entire set of factors that may influence the accuracy at this level. Fortunately, we initiated this process being aided with experience of successful development of a number of instruments that require similar technology and a comparable level of accuracy \citep{lator_cqg2004}. This experience comes with understanding various constituents of the error budget, expertize in developing appropriate instrument models; it is also supported by the extensive verification of the expected  performance with the set of instrumental test-beds (designed and built solely for this purpose) and existing flight hardware. 
Details of the LATOR error budget are still being developed (especially those of the second order) and will be published elsewhere, when fully analyzed. Here we discuss a preliminary design considerations that enable the desirable instrument performance. 

\subsection {Instrument Errors}

In our design we address two types of instrumental errors, namely the offset and scale errors. Thus, in some cases, when a measured value has a systematic offset of a few pm, there are may be instrumental errors that lead to further offset errors.  There are many sources of offset errors caused by imperfect optics or imperfectly aligned optics at the pm level; there also many sources for scale errors. We take a comfort in the fact that, for the space-based stellar interferometry, we have an ongoing technology program at JPL.  Not only this program has already demonstrated metrology accurate to a sub-pm level, but has also identified a number of the error sources and developed methods to either eliminate them or to minimize their effect at the required level.

A type of scalar error is introduced, for instance, by the laser frequency. Thus, in order to measure $\gamma$ to one part in  $10^{8}$ the laser frequency also must be stable to at least to $10^{-8}$ long term; the lower accuracy would result in a scale error. The measurement strategy adopted for LATOR would require the laser stability to only $\sim$1\% to achieve accuracy needed to measure the second order gravity effect. Absolute laser frequency must be known to $10^{-9}$ in order for the scaling error to be negligible. Similarly robust solutions were developed to address the effects of other known sources of scale errors. 

There is a considerable effort currently underway at JPL to evaluate a number of potential errors sources for LATOR, to understand their properties and establish methods to mitigate their contributions. (A careful strategy is needed to isolate the instrumental effects of the second order of smallness; however, our experience with SIM \citep{lator_cqg2004} is critical in helping us to properly capture their contribution in the instrument models.)  The work is ongoing, this is why the discussion below serves for illustration purposes only. We intend to publish the corresponding analysis and simulations in the subsequent publications.

\subsection{Optical Performance}
 
The laser interferometers use $\sim$2W lasers and $\sim$20 cm optics for transmitting the light between spacecraft. Solid state lasers with single frequency operation are readily available and are relatively inexpensive.   For SNR purposes we assume the lasers are ideal monochromatic sources (with $\lambda = 1.3~ \mu$m). For simplicity we assume the lengths being measured are 2AU = $3\times 10^8$ km. The beam spread is estimated as $\sim 1~\mu$m/20~cm = 5 $\mu$rad (1 arcsec). The beam at the receiver is $\sim$1,500 km in diameter, a 20 cm receiver will detect $1.71 \times 10^2$ photons/s assuming 50\% q.e. detectors. Given the properties of the CCD array it takes about 10 s to reach the desirable SNR of $\sim2000$ targeted for the detection of the second order effects. In other words, a 5 pm resolution needed for a measurement of the PPN parameter $\gamma$ to the accuracy of one part in $\sim10^{8}$ is possible with $\approx10$~s of integration.

As a result, the LATOR experiment will be capable of measuring the angle between the two spacecraft to $\sim0.05$~prad, which allows light deflection due to gravitational effects to be measured to one part in $10^8$. Measurements with this accuracy will lead to a better understanding of gravitational and relativistic physics. In particular, with LATOR, measurements of the first order gravitational deflection will be improved by a factor of 3,000. LATOR will also be capable of distinguishing between first order ($\propto G$) and second order ($\propto G^2$) effects. All effects, including the first and second order deflections, as well as the frame dragging component of gravitational deflection and the quadrupole deflection will be measured astrometrically.  
 
In our analysis we have considered various potential sources of systematic error. 
This information translates to the expected accuracy of determination of the differential interferometric delay of $\sim \pm5.4$ pm, which enables measurement of PPN parameter $\gamma$ to accuracy of  
%\begin{eqnarray}
$\gamma-1 = \pm 0.9  \times 10^{-8}.$
%\label{eq:accuracy}
%\end{eqnarray}
This expected instrumental accuracy is clearly a very significant improvement compared to other currently available techniques. This analysis serves as the strongest experimental motivation to  conduct the LATOR experiment.  

\subsection{\label{sec:expect_accuracy}Expected Measurement Accuracy}

Here we summarize our estimates of the expected accuracy in measurement of the relativistic parameters of interest.
The first order effect of light deflection in the solar gravity caused by the solar mass monopole is $\alpha_1=1.75$ arcsec; this value corresponds to an interferometric delay of $d\simeq b\alpha_1\approx0.85$~mm on a $b=100$~m baseline \citep{lator_cqg2004}. Using laser interferometry, we currently able to measure  distances with an accuracy (not just precision but accuracy) of $\leq$~1~pm. In principle, the 0.85 mm gravitational delay can be measured with $10^{-9}$ accuracy versus $10^{-5}$ available with current techniques. However, we use a conservative estimate  of 10 pm for the accuracy of the delay which would lead to a single measurement of $\gamma$ accurate to 1 part in $10^{8}$ (rather than 1 part in $10^{9}$), which would be already a factor of 3,000 accuracy improvement when compared to the recent Cassini result \citep{cassini_ber}. 

Furthermore, we have targeted an overall measurement accuracy of 10 pm per measurement, which for $b=100$~m this translates to the accuracy of 0.1 prad $\simeq 0.02 ~\mu$as. With 4 measurements per observation, this yields an accuracy of $\sim5.8\times 10^{-9}$ for the first order term.
The second order light deflection is approximately 1700 pm and with 10 pm accuracy and the adopted measurement strategy it could be measured with accuracy of $\sim2\times 10^{-3}$.  The frame dragging effect would be measured with $\sim 1\times10^{-2}$ accuracy and the solar quadrupole moment (using the theoretical value of the solar quadrupole moment $J_2\simeq10^{-7}$) can be modestly measured to 1 part in 20, all with respectable signal to noise ratios.

%************************************************

%************************************************
\section*{Conclusions}
\label{sec:conc}

LATOR mission is the 21st century version of Michelson-Morley experiment searching for a  cosmologically evolved scalar field in the solar system. This mission aims to carry out a test of the curvature of the solar system's gravity  field with an accuracy better than 1 part in 10$^{8}$. In spite of the previous space missions exploiting radio waves for tracking the spacecraft, this mission manifests an actual breakthrough in the relativistic gravity experiments as it allows to take full advantage of the optical techniques that recently became available.  LATOR has a number of advantages over techniques that use radio waves to measure gravitational light deflection. Thus, optical technologies allow low bandwidth telecommunications with the LATOR spacecraft. The use of the monochromatic light enables the observation of the spacecraft at the limb of the Sun. The use of narrowband filters, coronagraph optics and heterodyne detection will suppress background light to a level where the solar background is no longer the dominant noise source. The short wavelength allows much more efficient links with smaller apertures, thereby eliminating the need for a deployable antenna. Finally, the use of the ISS enables the test above the Earth's atmosphere -- the major source of astrometric noise for any ground based interferometer. This fact justifies LATOR as a space mission.

LATOR will lead to very robust advances in the tests of fundamental physics: it could discover a violation or extension of GR, or reveal the presence of an additional long range interaction in the physical law.  There are no analogs to the LATOR experiment; it is unique and is a natural culmination of solar system gravity experiments. 

%\vfill
%**************************************
%\acknowledgments
The work described here was carried out at the Jet Propulsion Laboratory, California Institute of Technology, under a contract with the National Aeronautics and Space Administration. 

% Bibliographic references with the natbib package:
% Parenthetical: \citep{Bai92} produces (Bailyn 1992).
% Textual: \citet{Bai95} produces Bailyn et al. (1995).
% An affix and part of a reference:
%   \citep[e.g.][Ch. 2]{Bar76}
%   produces (e.g. Barnes et al. 1976, Ch. 2).

%\end{harvard}
%***************************

\end{document}